# Agent-Based Model for River-Side Land-living

Portrait of Bandung Indonesian Cikapundung Park Case Study

Hokky Situngkir
[hs@compsoc.bandungfe.net]
Dept. Computational Sociology
Bandung Fe Institute

**Abstract**
A city park has been built from the organic urban settlement in the Cikapundung River, Bandung, Indonesia. While the aim for the development is the revitalization of the river for being unhealthy from the waste coming from the settlement. A study on how Indonesian people, in general, treating water source, like river, lake, and ocean is revisited. Throwing waste into the river has actually become paradox with the collective mental understanding about water among Indonesians. Two scenarios of agent-based simulation is presented, to see the dynamics of organic settlement and life of the city park after being opened for public. The simulation is delivered upon the imagery of landscape taken from the satellite and drone. While experience for presented problems gives insights, the computational social laboratory also awaits for further theoretical explorations and endeavors to sharpen good policymaking.

**Keywords**: agent-based model, computational social science, settlement, slum, river, water, waste management, indonesia

# 1. Introduction

Agent-based modeling has been one of important tools as a method to observe social phenomena [3[, due to the complex nature of social system [7]. Implementations have been there transgressing boundaries of social studies, from the social economic studies and the studies about law-enforcement [132], to the studies on epidemiology and economic and trade [11]. The social simulation through the agent based modeling have made it possible for social research delivered in the "computational laboratory" in which we can study the behavior of social system in an information theoretic fashion [12].

One interesting studies using the agent-based modeling is the emergence of settlement from the microstructure of housing due to the social and ecological interactions. The emerged settlement in any places comes from the abstraction on how people collectively react to the surrounding ecological aspects. How people think about mountains, sea, lake, rivers, climate, and so on, are the way the civilization evolved and reflected in how the settlements are there [*cf.* 2]. Here, computational simulation plays interesting role, for observer can see how the micro-properties of housing emerged into the macro patterns as discovered from the settlement.

As an archipelagic country, civilizations in Indonesia have great long history with water, sea, lake, and river. Water has major and important role for living, be it for daily domestic use and for irrigation and archipelagic activities. This is reflected in their face of settlement from different ethnic groups, different islands, and various religious and ways of life. Even in the modern era, it seems people seem to put the micro-social and collective thoughts in such ways emerging the similar pattern with those in rural areas. We would like to see this dynamical aspects computationally by seeing the settlements and how some ecological revitalizations policies delivered by the authorities in Bandung City with the case of study in Cikapundung River, the river flowing through the urban life.

The paper captures those aspects as materials to present an agent based modeling on the surface earth landscape and portrays. As a tool computationally putting time and space into dynamic observable presentation, agent based modeling promises reliable and comprehensive explanation about many complex social phenomena [1]. Ecological aspects is denoted by the layers with roles as inactive yet dynamical computational agents, and people with their emerged housings and settlements could be seen as animate and spatially changing ones [11].

# 2. Indonesian Social Living in Riverside

Since the ancient times. Indonesian ethnic groups have great appreciation for water within their ways of life. The diversity of ethnic groups in the archipelago gives variation to the similar emphasis on the importance of water source for social living. Among traditional Sundanese people in Western Java, water was named after "*kabuyutan*", literally means "ancestry", a devotion with magical and sacred narratives upon it [4]. As we related this to the ways Sundanese people living and establishing their settlements, we can see that water (whether in its form of lake or river, and sources) is something to do with conservation and protection for nature.

Among traditional Sundanese people, this is related to part of the concept named "*Tri Tangtu*", three important of one thing. Land use, they mentioned, should be functionally categorized as "*leuweung larangan*" (conservative function), "*leuweung tutupan*" (buffer or supporting function), and "*leuweung garapan*" (economic function) [9]. This is why, the descriptive portraits of Sundanese settlements always exhibit those three parts of land use [*cf.* 10]. It is interesting, as demonstrated in figure 1, the three parts is somehow emerged in other places in the archipelago. A sophisticated inquiries have been introduced in Bali, for instance, how the emphasis on the relation between the physical ecological aspect, like water, with the existing collective mental and spiritual abstraction made a kind of order and social structure in Balinese traditional irrigation system [6].

Simply, there are always land segregations in the land use, for function of water conservation, housing and settlements, and the land for economic production, for example the agriculture. There are always distancing boundaries between the three functions of land use that can generally discovered throughout the archipelago. They may come from different varies of traditional narratives and mental concepts, yet emerging similar patterns as we see the traditional settlements in Indonesia.

Moreover, this interesting facts are even emerged when we see how the organic settlements in some particular parts of the modern urban area. As Cikapundung riverbank becoming the place where people built houses since decades ago, the emerged settlement, in particular also exhibit the patterns of the three functional parts of land organically. However, since those from lower socio-economic class populated the houses within the settlement, it becomes slum, with not so healthy environment, especially as there is no integrated domestic waste management. In return, the river water has become dirty with domestic garbage. It becomes paradox, as the houses are built with some mental perspective of water as a place for conservation, but latter socio-dynamics let the water becomes ecologically unhealthy.

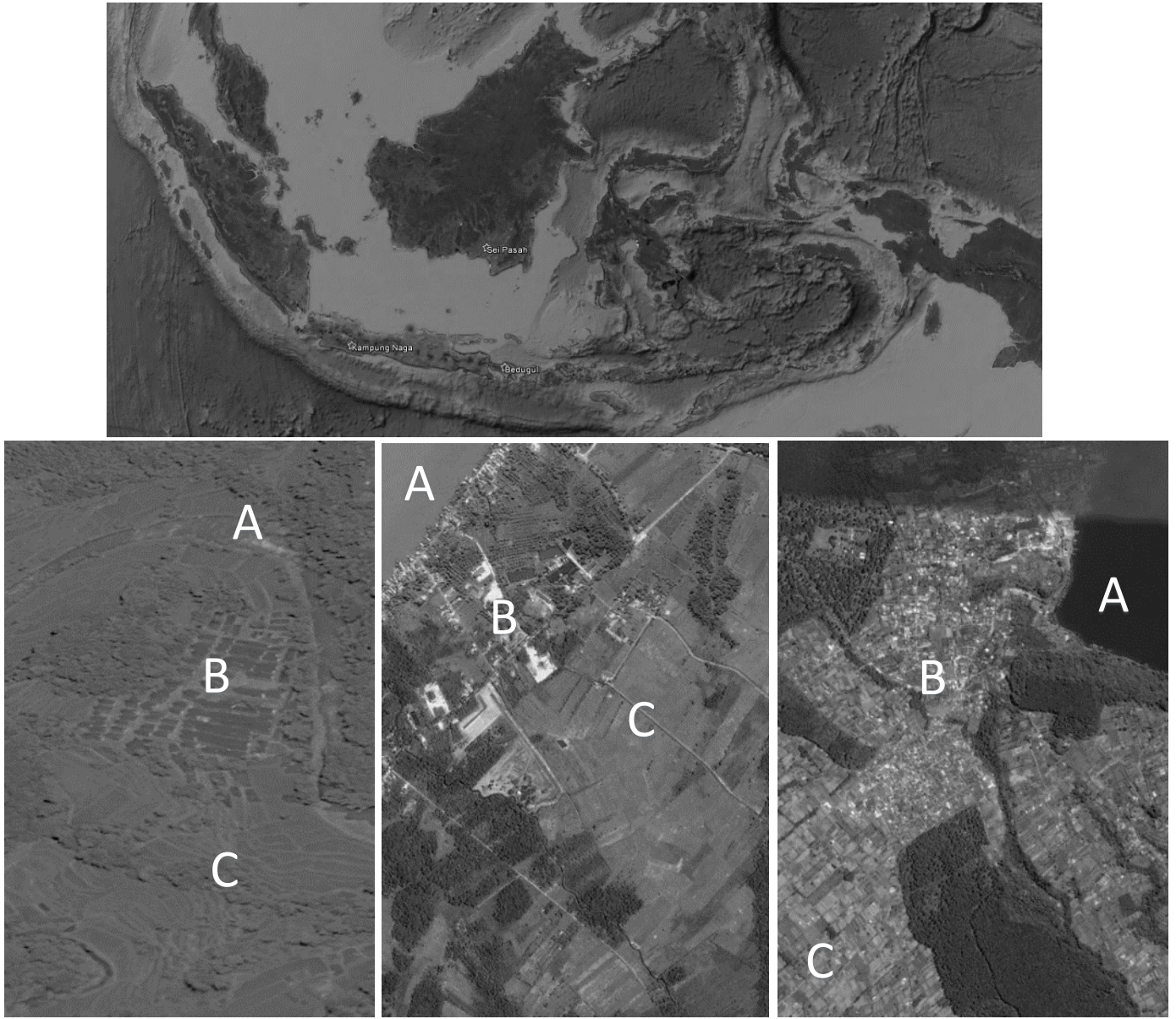

**Figure 1.** The organic settlement in Indonesia in the side of water-flow have particular micro-structural social process emerging the land use segregating the water-flows (A), the land for houses (B), and the agricultural land use (C). Exemplified in the picture the traditional settlement *Kampung Naga*, West Java (*left*), settlement in *Sei Pasah*, Borneo (*middle*), and *Bedugul* in Bali Island (*right*).

**3. Social Life Chronology in Cikapundung Riverside**
Bandung City in Indonesia, has been populated more and more dense every year. Places for building houses have become rarer in the past years. Riverbank, close to the river has become alternatives for those looking for spared land for shelter. Those people made buildings near the river, yet they followed some "rules" due to their collective knowledge about having houses within their customs and culture. The emerged settlement is as shown in figure 2 (*top*).

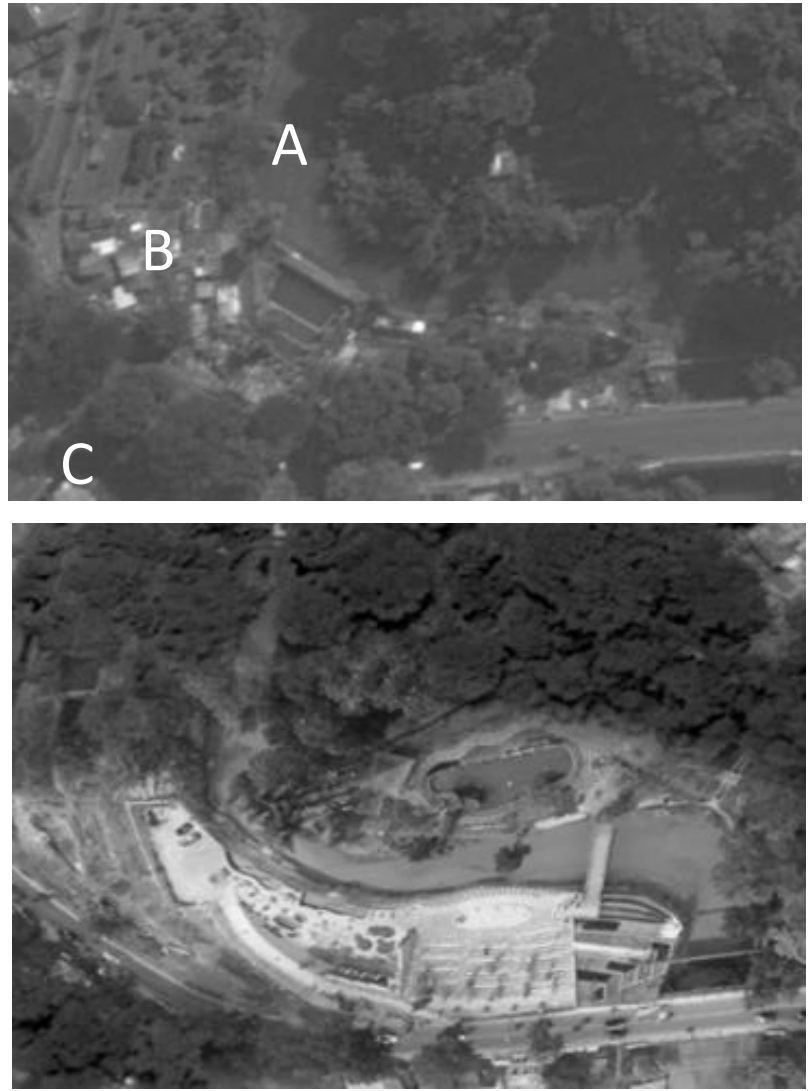

**Figure 2.** The urban organic settlement in Bandung city previously in the riverside of Cikapundung River was also emerged the spatial differentiation of land-use similar to the traditional one. The revitalization of the riverside by the authorities made the Cikapundung Terrace Park today.

Over the years, the settlement has become slum. There was no appropriate waste management, thus stack of domestic waste are there, and not long the river has become unhealthy. The brown water flows the Cikapundung river across the city while in some of many cases made the resistance of the flowing water, waiting to be flooded in the southern part of the city.

For the sake of the revitalization of the ecological landscape, the city government and the related national authorities have made the initiatives to move the settlements into more appropriate and healthy surroundings. The plan of city park is delivered. The city park has become green open public spaces since 2016, called "The Cikapundung Terace". The whole transformation of the riverbanks are shown in figure 2. The built city park has many "hot spots" for people to come and enjoy the open spaces. The scheme in figure 3 shows those hot spots.

However, the open public spaces still remain endangered by the indiscipline citizens enjoying their time in the park. The open space can not be conserved merely by physical development surrounding the river. However, there is some communities actively look after the park, and thus consequently the cleanliness of the river. The latter is there since the communities do business exploiting the existence of the river, for instance some water sports, rafting, and so on, of which success would depend upon the image of the clean river. The campaign of not wasting anything into the river has now buffered by community activities. For the ecological problem was rooted from social behavior, the effective social solution awaits.

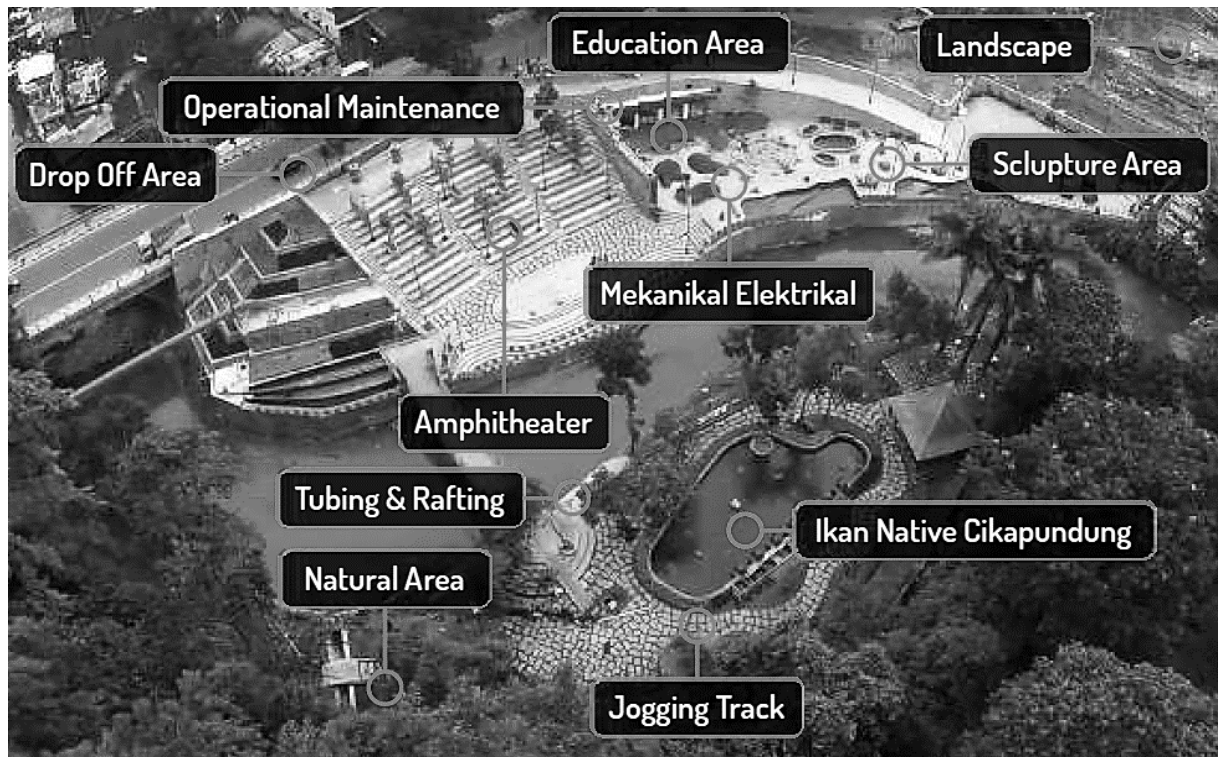

**Figure 3.** The "hot spots" within the Cikapundung Terrace recently.
(*courtessy of* BBWS Citarum)

**4. Artificial Society of Cikapundung Riverside**
The location of our case of study has witnessed at least two transformative ecological process. The first is the changing from open spaces into the establishment of the growing settlement. The second, is the development changing the settlement into the public spaces as city park. Two scenarios of dynamical process are proposed, both give the experience simulating people's behavior around the Cikapundung riverbank.

**4.1 Houses**
In order to understand how the organic settlement emerged from the housing process that has similarity with those shown in figure 1 and 2, we can implement algorithmic rule of building houses in an open landscape as adviced traditionally and widely known to people in the area. There are some suggestions that can be studied due to this issue. For instance, some open landscape near the riverside on which people suggested not to build house is [15],

- *Sri Madayung*. The land near and in between two streams of river, especially when the small stream in the left and the bigger one in the right side.
- *Talaga Kahudanan*. Near the exact land making river branching apart.
- *Si Bareubeu*. When the land is topographically below the streaming river (*katunjang ku cai*).
- Houses are traditionally also encouraged not to be built where seen highland or mountains put behind.

Those "rules" along with standards, like, building house closer one another to establish good neighborhood, and also closer to the part of the land where people earn their economy. Moreover, house should not too close to the flowing river in order to avoid flooding water in the rainy seasons, and in addition, some fact that people occasionally also use near river land for fishing, et cetera.

Those are general rules followed by the local people conventionally when they build houses in the settlement. The simplicity has emerged the complex behavior of settlement as we showed previously, and also we applied in our model in the first scenario of simulation.

**4.2. People**
On the lands are the people, interacting one another and with some features of the surrounding lands. On the land are some "landmark" and "hotspot" that randomly visited by artificial people. At time, $t$, people have utility function, $x_i(t)$ as,

$$x_i(t) = \frac{\sum_{neighbors} p_{neighbors}(t)}{N} - f_x(t)$$

where $p_{neighbors}$ denotes the "excitement" provided by the cells surrounding the location of the agent. Neighbors ($N$) we used here is the 8 square cells surrounding the agent. The crowded place and the unhealthy or dirty places ($\varepsilon$) with so many garbage surrounding would proportionaly (with some factor $\rho$),

$$f_x(t) = \left(\frac{\rho \sum_{neighbors} x_{neighbors}(t)}{N}\right) + \varepsilon$$

There is a gradual "diffusion-like" excitement over particular "hot spots" proportionally ($\mu$) among spots due to particular geo-location,

$$p = \frac{\mu(\sum_{neighbors} p_{neighbors})}{N}$$

The agents wander throughout the landscape visiting one "hot spots" from place to place with some recognition of where should step on based on whether or not some places can be walk through or not. Computationally speaking, this is artificially recognized from the image used as the landscape of the simulation, i.e.: the sattelite or image taken by drone.

**5. Simulation**
Two simulated scenarios are the landscape situation before and after the opening of the Cikapundung Terrace Park. Before the city park is built, the riverbank was an empty space. An algorithm to build houses of settlement near the river is implemented by using micro-properties as discussed in the previous section. Houses were tend to be located near the street as the road is their sources of income for urban works. They are tried to get further from the river if possible. The land close to the river were collectively made as "public" place where there was a place for fishing there and some field where children can gather or play in the day light.

From the houses built people live and going along with ecological features within the landscape, the delta, the trees, and the river. Houses generated domestic waste. Since it was a sort of slum in the city, there was little effort to for managing the waste. Many people there throw out their waste into the river. The river was becoming dirtier, and sometimes the surface of the water was high, even though the location was not flooded by the river for the topography was quite higher towards the main road. The figure of the situated simulation is shown in figure 5 (*top*).

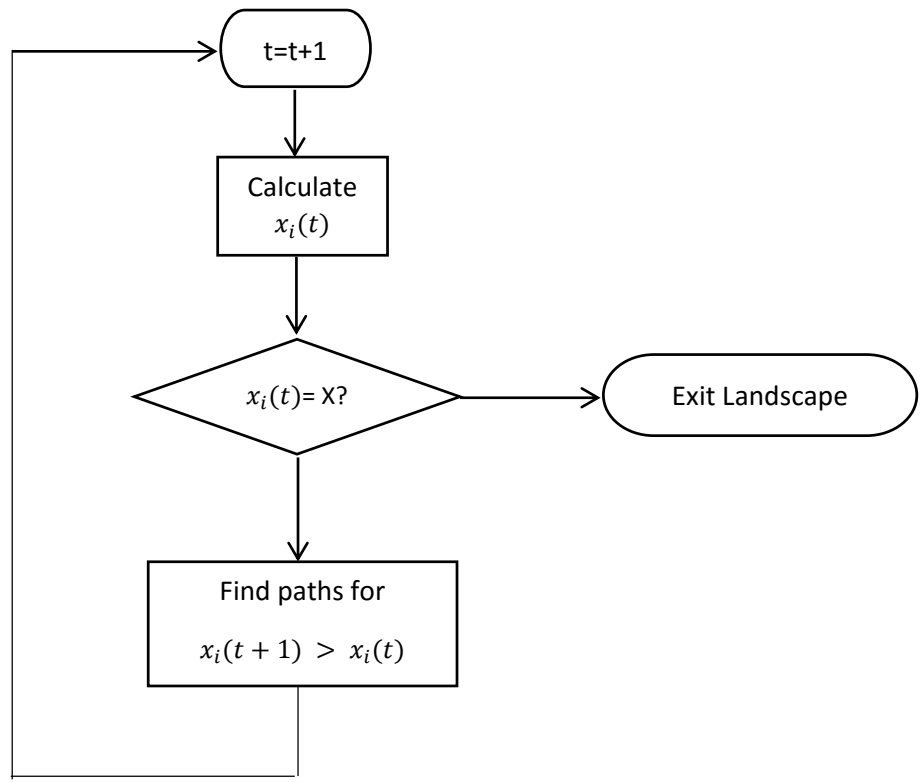

**Figure 4.** The flow chart of wandering agents within the whole landscapes.

In the second scenario of the simulation, all the houses were demolished. There were no houses for the land was made into open space. The only people within the landscape is the citizens coming to the public space for leisure and get fresh open air. People wanders throughout the park from hotspots to hotspots. The hotspots in the park are shown in figure 3. As the utility function of the people has been discussed in the previous section, the way people wander [*cf.* 5, 8] throughout the landscape is depicted in the flowchart in figure 4.

The simulation is in discrete time-space based on map [*cf.* 1] in Netlogo [16], and the artificial people wanders in based on how they locally perceive the artificial world. The visitors of the park have different kinds of waste or rubbish. Bottles of drink, wrappings of foods, newspapers, and more are more likely to be the garbage left by the visitors, while previously the problems were more likely to be domestic waste, which were much more complex. However, the potential to counter revitalization of the river were still there.

In reality, an organized community, named "Komunitas Cikapundung", supports the city park. This community tried to keep the park clean by doing educational campaign for clean river and showing where to put garbage. They also let the visitors to enjoy the clean river by providing some water activities in the river. This has actually made a great improvement for the river to stay clean. People become reluctant to throw out rubbish into the water.

In our simulation, there are agents play role as the community member, wandering around not to let any garbage wasted within the park. In this simulated scenario, people will not throw out anything in the park. However, some inevitable opportunistic visitors still do the leftovers when there are no other people warn them not to do so. This city park scenario is demonstrated in the bottom panel of figure 5.

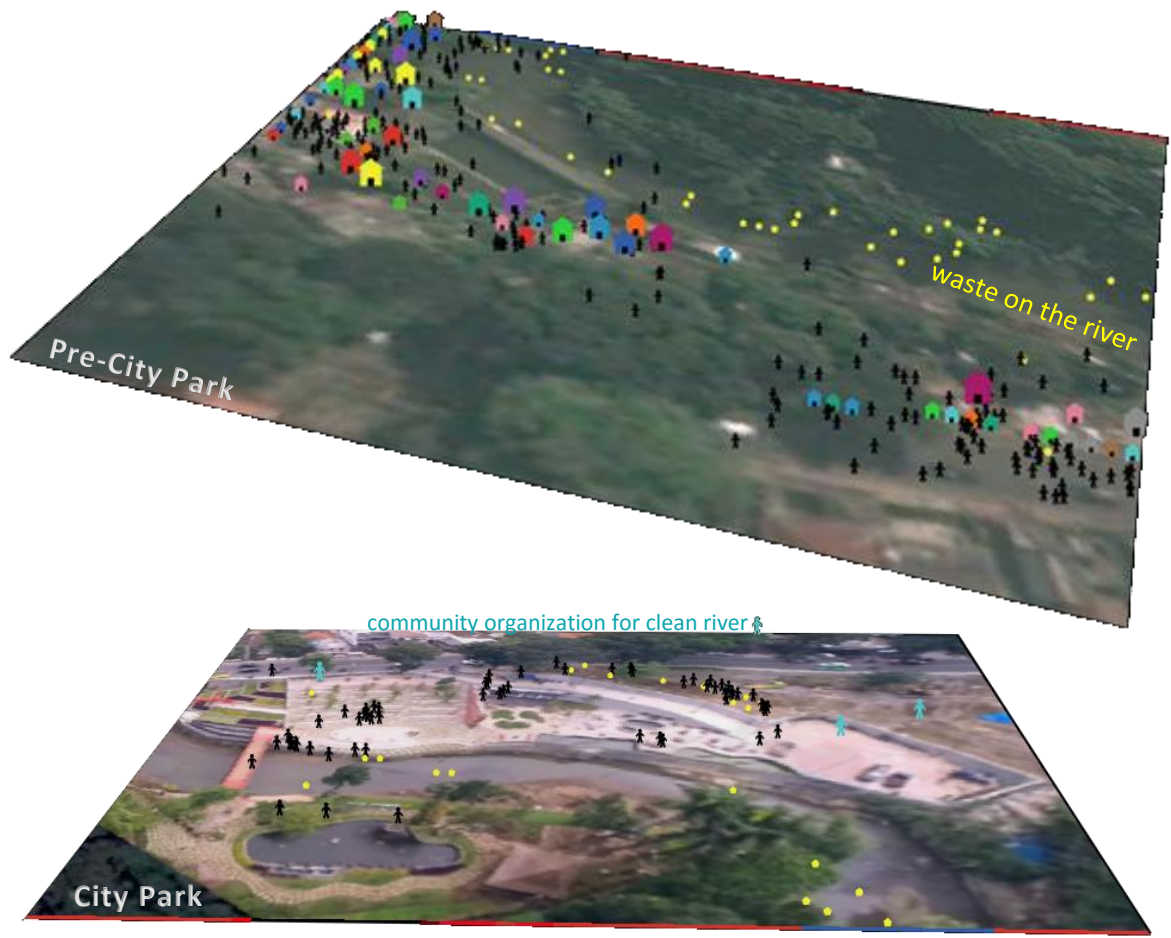

**Figure 5.** Two simulation landscapes, scenario for pre city park and after the opening of the *Teras Cikapundung*.

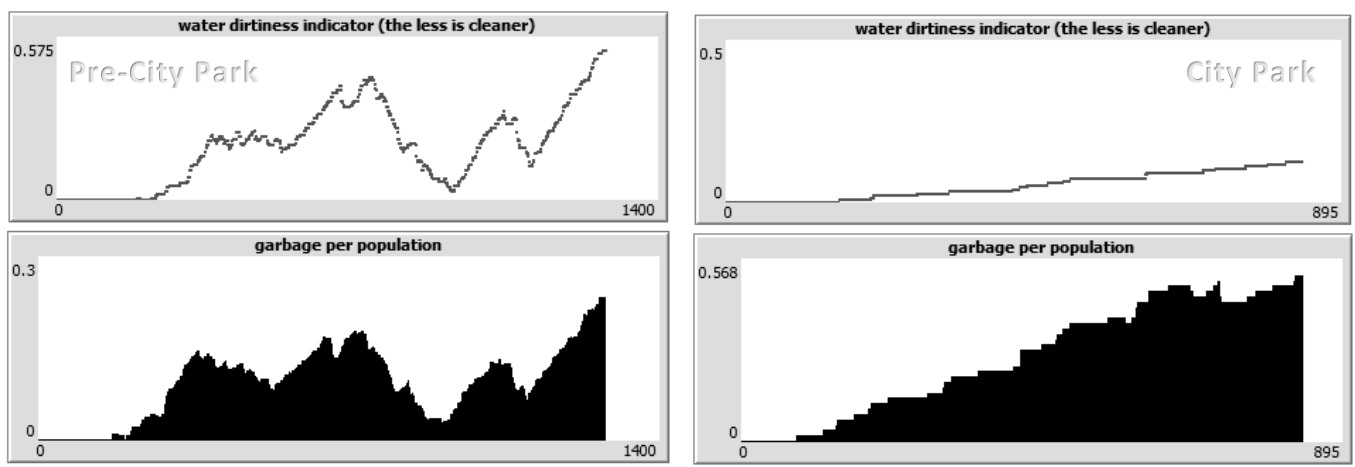

**Figure 6.** The panels comparing the waste situation before and after the opening of the city park.

We discussed about settlement and interaction among artificial citizens and park visitors yet our focus in the simulation for all scenarios are actually the revitalization of the some particular parts of Cikapundung River. The result of simulation is shown in figure 6. It compares the waste situations before and after the opening of the Cikapundung Terrace Park. There is definitely a drastic change in the waste thrown into the river for most of the citizens there were moved away from the riverbank. However, as discussed previously, the tendency to waste among the populations does not change at all.

People will always tend to waste, and without any action to prevent it, the stacked garbage will always be there. In the figure, we show the experiment that even the garbage per population is slower in the slum; the rate of the index of dirtiness of the river water is drastically damped after the park. This is one of the experiment result that can be managed and presented within the artificial society in our simulation.

**6. Concluding Remarks**
It is an emerged paradox that traditionally water, in its physical existence like river, lake, and ocean, is a very important thing in the abstraction of life among Indonesian people, but people keep throwing waste into the water. It is possible that it came from the hardship of urban living that distract of things people have in their mental mind.

The computational simulation with the two scenarios has demonstrated it and let observer to experience the factual situation. The simulation runs on the supervised distinction for features of earth imagery (from the satellite or drone) landscape to see artificial social agents interact with them, as well as interacting to one another dynamically. The simulation, since it is presented on the photo imagery from the above might give interesting experience looking at the system. Moreover, the simulation has depicts how the transformation of land use from the slums into city park has given positive result for the revitalization of river and the public community awareness and actions for saving the ecology.

The agent based modeling has shown its interesting role on explaining phenomena by giving experience to the observer. From such simulation, experiments can be delivered for the sake of social theoretical exploration, as well as the sharpening policy-making.

**Acknowledgement**
The author thanks Yayat Yuliana from BBWS Citarum for sharing his experience led the revitalization of the Cikapundung Terrace, IGW Samsi Gunarta, Monika Raharti, and crews in the Center for Young Scientists for initiating the discussions about the subject matter, and Guntur Purba from Bandung Fe Institute, for assistance in which period the work is delivered. All faults remain author's.